# Advanced structural characterization of single-walled carbon nanotubes with 4D-STEM

Antonin LOUISET[1], Daniel FÖRSTER[2], Vincent JOURDAIN[3], Saïd TAHIR[3], Nicola VIGANO[1], Jean-Luc Rouvière[1], Christophe BICHARA[4], and Hanako OKUNO[1].

[1]IRIG-MEM, CEA, Université Grenoble Alpes, Grenoble, France.

[2]Interfaces, Confinement, Matériaux et Nanostructures, ICMN, Université d'Orléans, CNRS, Orléans, France.

[3]Laboratoire Charles Coulomb, CNRS, Université de Montpellier, Montpellier, France.

[4]CINaM, CNRS, Université Aix-Marseille, Marseille, France.

*Author to whom correspondence should be addressed: hanako.okuno@cea.fr*

**Abstract**

Single wall carbon nanotubes (SWCNT) exhibit remarkable optical and electrical properties making them one of the most promising materials for next generation electronic and optoelectronic devices. Their electronic properties strongly depend on their chirality, i.e., their structural configuration, as well as on the presence and nature of atomic defects. Currently, the lack of versatile and efficient structural characterization techniques limits SWCNT applications. Here, we report how four-dimensional scanning transmission electron microscopy (4D-STEM) can address critical challenges in SWCNT structural analysis. Using modern fast pixelated electron detectors, we were able to acquire rapidly a large number of low noise electron diffraction patterns of SWCNTs. The resulting 4D-STEM data allow to precisely determine the local chirality of multiple nanotubes at once, with limited electron dose (down to $1.75 \times 10^3$ e-/Å²) and nanometric spatial resolution (down to 3.1 nm). We also show how this approach enables to track the chirality along a single nanotube, while giving access to the strain distribution. Then, we report how 4D-STEM data enable to reconstruct high-resolution images with electron ptychography. With this second approach, structural information can be obtained with atomic scale spatial resolution allowing atomic defect imaging. Finally, we investigate how multi-slice electron ptychography could provide even further insight on nanotube defect structures thanks to its close to 3D imaging capabilities at atomic resolution.

# 1. Introduction

Single-walled carbon nanotubes (SWCNTs) exhibit remarkable electronic properties (high mobility, small capacitance) and have a wide range of current and potential applications in electronics [1,2], optoelectronics and photonics [3–5], bioimaging [6,7] and more. The electronic structure of SWCNTs is governed by its atomic arrangement imposed by the tubular form, enabling tunable optic and electronic properties. In particular, chirality is a unique structural parameter of SWCNTs playing a decisive role in the determination of their structure-dependent electronic properties. The atomic structure of SWCNT can be described as graphene sheets rolled into cylindrical form. Since the sheet can be rolled in countless ways, they exhibit a diverse range of helical structures. The helical configuration of a SWCNT (chirality) is defined by its chiral indices which are expressed by two integers (n, m). They set the nanotube chiral vector $\boldsymbol{C_h} = n\boldsymbol{a_1} + m\boldsymbol{a_2}$, with $\boldsymbol{a_1}$ and $\boldsymbol{a_2}$ being the unit vectors of the graphene honeycomb lattice and $|\boldsymbol{C_h}|$ being the perimeter of the tube. A SWCNT is metallic when $n - m = 3l$, with $l$ an integer, whereas semiconducting when $n - m = 3l \pm 1$, with the bandgap, in a first approximation, scaling with the inverse of the tube diameter [8]. The performance of SWCNT-based devices thus fundamentally depends on the ability to control their chirality.

For this purpose, significant efforts have been devoted to the development of both chiral-selective growth and effective post-growth sorting methods for large-scale applications. Today, the best candidate to achieve chiral-selective growth is the catalytic chemical vapor deposition (C-CVD) method [9,10]. It is capable of reaching 97% chiral purity [11] and even 99.9% semiconducting purity when combined with electro-renucleation [12]. However, this is still far from the chiral purity that would be required for modern transistor applications: > 99.9999 % of semiconducting tubes [1]. Most of the time, SWCNT-based devices require chirality-pure nanotubes which involve an additional sorting step after the growth [6,10]. In addition, recent studies showed that chirality mutations, *i.e.* variations of chiral indices along an individual nanotube, arise during nanotube growth, or can be induced afterwards [13–15]. These mutations can be engineered, opening new possibilities for nanotube electronics and nanoscale quantum devices [16]. However, the origins of chiral selectivity and chirality mutations as well as the mechanism underlaying the formation of the associated atomic defects remain controversial. As demonstrated in graphene, growth related atomic defects such as grain boundaries are expected to strongly influence the optical, chemical, and electronic properties of SWCNTs. In graphene, the detailed structures and formation mechanisms of such defects are relatively well understood, supported by numerous reliable experimental investigations and associated theoretical studies [17]. In contrast, the understanding of the structure and formation mechanism of intramolecular junctions, atomic defects formed between regions with two different chiralities along a SWCNT, is still mostly theoretical [18–20]. Only a few studies provided reliable characterization of their structure [21] and some results remain debatable [14]. This is due to the challenges associated with structural characterization of SWCNTs, arising from their peculiar structure: a high aspect ratio with ~1 nm diameter, lengths often exceeding a tenth of a micron, and a three-dimensional tubular structure. Therefore, robust analytical techniques capable of precisely determining local chirality and the structure of associated atomic defects in individual SWCNTs are still necessary to gain a deeper understanding of the fundamental growth mechanisms in C-CVD and further to achieve highly controlled electronic properties of SWCNTs, and which are crucial for next-generation carbon nanotube applications [10,16].

The most common chirality assignment techniques, such as Resonant Raman Spectroscopy (RRS), Photoluminescence spectroscopy (PL) and Rayleigh Scattering Spectroscopy (RSS), are non-destructive, and allows rapid probing of a large number of nanotubes which enables efficient statistical analysis [10]. However, these methods are prone to artifacts and intrinsic biases, leading to errors in chiral index measurements or even the non-detection of certain chiralities [10,22–24]. To ensure robust chirality determination, optical methods often need to be combined. Furthermore, their spatial resolution is far above the nanometer scale, preventing the separate analysis of adjacent tubes or the high-resolution tracking of chirality mutations.

One of the best alternatives to optical techniques is Transmission Electron Microscopy (TEM). To resolve the helical structure of nanotubes, researchers first used Nano Beam Electron Diffraction (NBED) [25], which is now a standard experiment for TEM equipment. Early works showed that there is an analytical solution to describe the structure factor of a (*n*, *m*) SWCNT, and therefore its diffraction pattern [26]. This allowed Jiang *et al.* to develop, a method for unambiguous chirality determination from these patterns, the "intrinsic layer-line spacing" (ILLS) [27], which is still in use today [28]. While this method is robust to assign the global chiral indices, it can only be applied to isolated and suspended SWCNTs, which requires dedicated sample preparation. Due to the small scattering power of SWCNTs, NBED requires large electron doses, typically between $10^3$ to $10^5$ e$^-$.Å$^{-2}$ [29,30], to reach sufficient counts or Signal to Noise Ratio (SNR). Most of the time, this leads to a long exposure time when recording diffraction patterns, from a few seconds to more than 10 [30–32], which limits the possibility to perform many acquisitions in a row. Furthermore, a single diffraction pattern only contains reciprocal space information, so local structural changes along a tube cannot be identified. Finally, due to the relatively large diameter of the electron beam (tens of nm [26]), it can only be applied to isolated nanotubes. Later, the advent of Aberration-Corrected TEM (AC-TEM) has enabled the use of direct imaging for chirality determination. Chiral indices can be measured from High-Resolution TEM (HR-TEM) images, either by analyzing the image Fourier transform [33] or directly from the moiré patterns appearing in the real space image using deep learning algorithms [34]. Furthermore, atomically resolved images offer the opportunity to study atomic defects, although, interpretation of HR-TEM images of SWCNTs heavily relies on simulation to resolve the defect's structure [35]. While this method presents multiple advantages compare to NBED, it is rarely applicable to free-standing nanotubes which tend to vibrate under electron beam irradiation. A workaround consists in transferring SWCNT on a graphene layer to act as a low contrast mechanical support [33,36] but this additional step complicates the sample preparation and images are harder to interpret.

Here, we present two complementary approaches based on 4-Dimensional Scanning TEM (4D-STEM), to overcome the current limitations in chirality determination with TEM. 4D-STEM is a powerful technique that records a diffraction pattern at each probe position during a STEM scan, resulting in a dataset with both real and reciprocal space information [37]. In the first implementation, a small convergence angle is used, which enables quantitative analysis of diffracted intensities to extract chirality maps of SWCNTs. This NBED-like approach provides the best sensitivity to chirality while improving the spatial resolution compared to standard NBED and introduces an error metric to assess the measurement quality. This type of 4D-STEM dataset also contains information about the strain field in the material. In this work, we apply this 4D-STEM approach to map strain, specifically the radial strain component, since the diameter of SWCNTs directly influences their electronic and optical properties [8].

Secondly, a large convergence angle is used to improve spatial resolution. This high-resolution 4D-STEM configuration gives access to advanced computational image reconstruction methods, in particular electron ptychography. This technique reconstructs the complex transmission function of the sample and provides phase contrast imaging with sub-angstrom resolution [38,39], enabling the visualization of individual carbon atoms and fine structural details in SWCNTs [40–42]. Even with outstanding spatial resolution, resolving the structure of defects in SWCNTs remains challenging due to the projected moiré patterns intrinsic to tubular structures. To address this issue, we explore multi-slice electron ptychography, a more advanced phase retrieval technique that offers not only higher spatial resolution than conventional (single-slice) ptychography but also enables three-dimensional imaging through numerical depth sectioning [43–45]. Finally, we discuss its potential for directly visualizing atomic defects in SWCNTs.

## 2. Methods

### 2.1. Sample preparation

The nanotubes were synthesized by CCVD on homemade TEM grids. The latter were fabricated by optical lithography and plasma etching (RIE) from multiframe arrays purchased from Oxford Instruments. The catalyst (1 Å cobalt) was deposited on one half of the grids by vacuum thermal evaporation. The grids were heat-treated in air, and then introduced into the CVD reactor airlock. After purging the reactor to remove oxygen and water, the furnace was heated to 850 °C. After a 5-minute temperature stabilization, the sample was rapidly introduced into the heated zone and the synthesis continued for 30 minutes. Ethanol was used as the carbon precursor, carried by a flow of argon (66.6 sccm) through a bubbler maintained at 0 °C. The total flow was supplemented by another line of argon at 832.7 sccm and dihydrogen at 189.0 sccm. After 30 minutes, the samples were quickly removed from the heating zone.

### 2.2. Nanotube molecular dynamic simulations

To evaluate the performance of electron ptychography, artificial 4D-STEM datasets were generated from defective nanotube structures. Realistic atomic configurations were obtained through molecular dynamics (MD) simulations computed with the LAMMPS package [46]. These MD simulations are based on the Tersoff potential [47] known to be a reliable description of carbonaceous structures [48,49]. The simulations start off at 6500 K to induce the formation of topological defects. The system is then rapidly cooled down to room temperature. The details of the simulation protocol are published elsewhere [34]. Once equilibrium is reached, 16 snapshots at 300 K are extracted as structural models for the 4D-STEM simulations (see next section).

### 2.3. 4D-STEM experiments and simulations

4D-STEM experiments were performed on a Titan Ultimate microscope equipped with an X-FEG gun and 2 spherical aberration correctors for probe and image modes. The incident electron beam energy was set at 80 keV to prevent the tubes from being damaged. For chirality and strain mapping, the convergence semi-angle of the beam (α) was set at 0.8 ± 0.05 mrad (small angle configuration) and the real space pixel size was set at 1.04 nm with an exposure time of 50 ms per pixel. For the electron ptychography experiment, the convergence semi-angle of the beam was set at 26.1 ± 0.05 mrad (large angle configuration) and the exposure time per scan position was 2 ms. The real space pixel size was 0.76 Å and the reciprocal space pixel size was 0.0152 Å$^{-1}$. The electron dose per area was approximately $10^6$ e$^-$.Å$^2$. The STEM probe was in-focus to directly compare the reconstruction with the Annular Dark Field (ADF) image. All datasets were recorded on a Medipix3 camera composed of 256x256 pixels. This hybrid pixel direct electron detector offers several advantages, including low noise and a large dynamic range. This wide range eliminates the need for a beam stop and prevents saturation near the direct beam. It is also crucial for electron ptychography, where both strong and weak signals must be captured simultaneously.

Diffraction pattern and 4D-STEM dataset simulations were carried out with the abTEM package [50] within the multi-slice formalism. To simulate small convergence angle experiments, parameters were set at 0.142 Å for the lateral real space sampling, with a slice thickness of 1.42 Å. For the large convergence angles, parameters were set at 0.08875 Å for the lateral real space sampling, with a slice thickness of 0.71 Å. To emulate thermal vibrations at room temperature during the experiment, we use the frozen phonon approximation using 16 the molecular dynamics snapshots as input structures (only for large convergence angles). Simulations were carried out to emulate different experimental conditions. A first one to match our experimental conditions, meaning: 80 kV with a convergence angle of 26.1 mrad considering an energy spread of 1.2 eV, typical of a X-FEG gun. A second one to represent a more coherent cold-FEG (CFEG) gun with an energy spread of 0.4 eV at 200 kV with the same convergence angle. For both large convergence angle simulations, a focal spread δ was introduced to consider the energy spread of the electron beam and the chromatic aberration of the microscope. δ values were computed using the formula

$$\delta = C_c \frac{\Delta E}{E}, \qquad (1)$$

where $C_c$ is the chromatic aberration, $E$ and $\Delta E$ are the energy and energy spread of the electron beam, respectively. We used 7 probe wave functions using the weighted incoherent integral method to emulate the focal spread [51]. Then, Poisson noise was added to the final 4D datasets corresponding to an electron dose per area of $10^6$ e$^-$.Å$^2$. For all configurations, we used the finite projection approximation for the calculation of the electrostatic potential in each slice.

### 2.4. Chirality determination
#### 2.4.1. Intrinsic layer-line spacing method

To measure the chirality from small convergence angle diffraction pattern, we used the ILLS method described in [27]. We briefly summarize the method in what follows but first we need to introduce the specificities of the electron diffraction pattern of a SWCNT. When an electron probe with a small convergence angle is transmitted through a nanotube, as represented in **Figure 1**(a), the resulting diffraction pattern is composed of diffraction spots arranged in a set of lines which originate from the helical symmetry of the nanotube and are characteristic of its chirality. **Figure 1**(b), shows a typical experimental diffraction pattern of a chiral tube ($m > 0$ and $n \neq m$). The most intense line at the center is the equatorial line ($L_0$), then there are three $1^{st}$-order lines, ($L_1, L_2, L_3$), and finally three $2^{nd}$-order lines, ($L_4, L_5, L_6$). They are all oriented along the tube radial direction, $\mathbf{k_r}$.

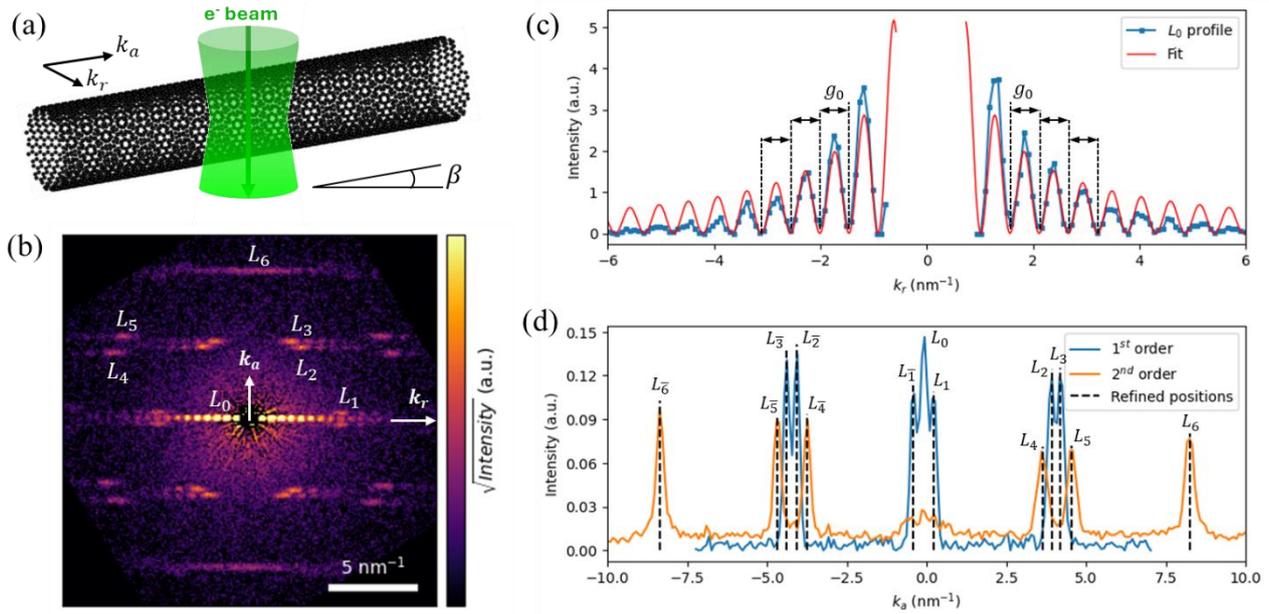

**Figure 1** - Chirality determination from a nanotube diffraction pattern. (a) schematic showing the orientation of a nanotube relative to the incident convergent electron beam. (b) experimental diffraction pattern of a SWCNT, without the direct beam and plotted with a square root intensity scale. (c) Intensity profile of the equatorial line extracted from (b), blue squares connected by a blue line and the associated fitted $0^{th}$-order Bessel function, red line. (d) $k_a$ intensity profiles integrated over $k_r$ considering only the $1^{st}$ and $2^{nd}$ order lines, blue and orange lines, respectively, and refined peaks positions marked by black dashed lines.

The ILLS method comprises two steps: first, the tube diameter, $d_0$, is measured by fitting the $L_0$ intensity profile, and then, the tube geometry is determined by measuring the spacing between each non-equatorial layer-line, $L_i$, and $L_0$. By combining all this information, it becomes possible to determine the nanotube's chiral indices while simultaneously correcting the tilt angle, $\beta$, between the tube axis and the direction perpendicular to the electron beam propagation direction [27] (see **Figure 1**(a)), without relying on diffraction calibration.

Let us start by measuring the tube diameter by extracting and fitting the intensity profile of the equatorial line along $k_r$: $I_{L_0}(k_r)$. Within the framework of kinematical diffraction theory, we can show that $I_{L_0}(k_r)$ is proportional to the squared modulus of a Bessel function [26,27], such as:

$$I_{L_0}(k_r) = C|J_0(\pi d_0 k_r)|^2, \qquad (2)$$

where $C$ is a strictly positive real constant, $d_0$ is the diameter of the observed tube and $J_0$ is the $0^{th}$-order Bessel function (see Eq. (A2) in Appendix for the definition of $J_0$). The profile extracted from **Figure 1**(b) is plotted in **Figure 1**(c) (blue squares connected by a blue line) with its corresponding fit (red line). Note that the distances between the roots of $J_0(\pi d_0 k_r)$, *i.e.*, distances between the $I_{L_0}(k_r)$ minima, are all equal to $g_0 = \frac{1}{d_0}$ [27]. Therefore, the refined $d_0$ value is reliable as long as the positions of the maxima and minima of the experimental data and fitted function are the same, even if there is a large amplitude discrepancy between the two. In previous studies, the $I_{L_0}$ fitting procedure was performed using an approximation of the $0^{th}$-order Bessel function [27,52]. Here, we implemented a non-linear least square fitting procedure using the SciPy software [53], which includes direct computation of Bessel functions without approximations [54]. Because small variations in $d_0$ induce large changes in $I_{L_0}$ periodicity and amplitude, we chose to manually pass a first guess for C and $d_0$ values prior to the fitting process to improve its robustness.

Then we measure the layer-line spacings, $G_i, (i = 1, 2, \ldots, 6)$, which are define as the reciprocal space distances between $L_i$ and $L_0$, along the tube axial direction $k_a$. This is rather straightforward as it consists in getting the overall intensity profile along the tube axial direction $k_a$, by integrating the diffracted intensity over $k_r$ and measuring the $k_a$ coordinates of the peaks. **Figure 1**(d), shows the integrated intensity profiles for the $1^{st}$ and $2^{nd}$ order lines, (blue and orange line respectively). They are composed of a series of peaks, each one corresponding to one layer-line. The black dashed lines represent the peak positions refined using the SciPy toolkits. **Figure 1**(d) lines with a positive $k_a$ component are labelled $L_i$ and lines with a negative one are labelled $L_{\bar{i}}$.

$G_i$ contain information about the $\frac{n}{m}$ ratio [26] but they depend on the tilt angle, $\beta$. To determine the absolute values of chiral indices, we need to compute the unit less intrinsic layer-line spacings define as:

$$\xi_i = d_0 G_i = \frac{G_i}{g_0}. \qquad (3)$$

Each $\xi_i$ are defined as a linear combination of the chiral indices (see Appendix). However, the quantities measured experimentally, $\xi_i^\beta$, also depend on $\beta$. They are related to the intrinsic values following the relation:

$$\cos\beta = \frac{\xi_i}{\xi_i^\beta}. \tag{4}$$

To determine the chirality of the tube, we need to solve for 3 unknowns: $n$, $m$ and $\beta$. To do so, we have 12 available equations: 6 for all $\xi_i$ (Eqs. (A3)), and 6 times the $\cos\beta$ relation (Eq. (4)). After solving this system, we obtain real values for the chiral indices: $n^\beta$ and $m^\beta$. The actual chiral indices are obtained by recovering only the integer part of the measured values, such as:

$$n = \lfloor n^\beta \rfloor, \ m = \lfloor m^\beta \rfloor, \tag{5}$$

where $\lfloor \cdot \rfloor$ is the floor function. Some additional corrections are required for large $\beta$ values and can be found in [27]. In addition of providing robust chirality measurement, this method does not require any prior calibration of the reciprocal space pixel size because $\xi_i$ are unitless. In the following, we will also frequently use calibrated data to evaluate the accuracy of the method for measuring diameters.

### 2.4.2. Relative and absolute error estimation

The hybrid pixelated detector used in this study only contains a few numbers of pixels compare to conventional CMOS or CCD cameras. As a result, a one-pixel error in the line position measurement can significantly alter the chirality determination. Therefore, it is crucial to estimate the potential errors in the chirality measurement. These can be directly estimated by the fitting procedure itself using the covariance matrix. A standard deviation error, $\sigma$, for each parameter is obtained by taking the square root of the diagonal elements of this matrix. Therefore, when fitting the $I_{L_0}$ intensity profile, a $3\sigma_{d_0}$ confidence interval can be directly recovered. However, in the case of layer-line spacings, the peak position refinement has to be done with a limited number of pixels (between 3 to 11 pixels per peak, depending on the dataset). This reduced number of data points results in inaccuracies in the estimation of the covariance matrix. Fortunately, the errors in the layer-line spacings can be assessed using an alternative method.

As detailed in the previous section, multiple $\xi_i^\beta$ can be combined to determine the tilt-affected chiral indices: $n^\beta$ and $m^\beta$. Here, we chose to measure the indices with three different layer-line spacing combinations: a first one by combing $\xi_2^\beta$ and $\xi_3^\beta$, a second one by combing $\xi_3^\beta$ and $\xi_6^\beta$ and a last one by combing $\xi_4^\beta$ and $\xi_5^\beta$ (see Eqs. (A3) in Appendix). To reduce possible errors, the 3 $(n^\beta, m^\beta)$ values are averaged to give the final $(n,m)$ results. Then, the intrinsic $\xi_i$ are computed which allows to calculate six different values for the tilt angle, using Eq. (4) and rearranging such as

$$\beta_i = \cos^{-1}\left(\frac{\xi_i}{\xi_i^\beta}\right). \tag{6}$$

Once again, the final $\beta$ value is obtained by averaging all six $\beta_i$ values. The only piece of information not used for chirality determination is $\xi_1^\beta$. Therefore, it can be used to estimate the residual error. We can compute a theoretical value for $\xi_1^\beta$ using the relation

$$\xi_1^\beta = \frac{n^\beta - m^\beta}{\sqrt{3}\pi}, \tag{7}$$

(see Eqs. (A3) in Appendix). Absolute and relative errors can be calculated by comparing the measured $\xi_1^\beta$ with the right term in Eq. (7). The relative error, $\varepsilon_{rel}$, is expressed as

$$\varepsilon_{rel} = \frac{\sqrt{3}\pi\xi_1^\beta - n^\beta + m^\beta}{n^\beta - m^\beta}. \tag{8}$$

Because $\xi_i^\beta$ are unitless, they are not well-suited for estimating absolute errors, are not well-suited for estimating absolute errors, especially when expressing them as a number of pixels on the detector in reciprocal space. To do so, it is more convenient to scale the absolute error by $\frac{1}{d_0}$, which allows it to be represented in the same units as the layer-line spacing $D_i$. By subtracting the two terms in Eq. (7) and using Eq. (3), we obtain the following for the absolute error, $\varepsilon_{abs}$, expressed in reciprocal pixel scale:

$$\varepsilon_{abs} = \frac{\sqrt{3}\pi\xi_1^\beta - n^\beta + m^\beta}{\sqrt{3}\pi d_0}. \tag{9}$$

Note that for armchair and zigzag tubes, this error cannot be estimated because there are not enough layer-lines to get information redundancy. However, these structural configurations can be easily identified due to their characteristic diffraction signature, and chirality assignment errors arise only from uncertainties in diameter measurement.

## 2.5. *Chirality mapping*

To generate a chirality map, an ADF image of the captured area is reconstructed from the 4D-STEM dataset using a virtual detector. In the ADF image shown in the top of **Figure 2**(a), we see a typical nanotube bundle separation with its characteristic Y-shape [55–57], composed of two SWCNTs, NT1 (left) and NT2 (right). There are also regions with a brighter contrast in this image highlighted by the yellow arrows. They correspond to residual contaminants (mostly carbon) which is common in CVD grown SWCNTs [22,58,59]. To determine the chirality of each nanotube, close to the junction, we segment the ADF image to isolate regions of interest (ROI), *i.e.*, individual tube without contamination. Then, we extract the diffraction corresponding to the ROI of NT1 (blue area in **Figure 2**(a)), and NT2 (red area in **Figure 2**(a)). These "sub-datasets" are represented at the bottom of **Figure 2**(a), for NT1 (left) and NT2 (right).

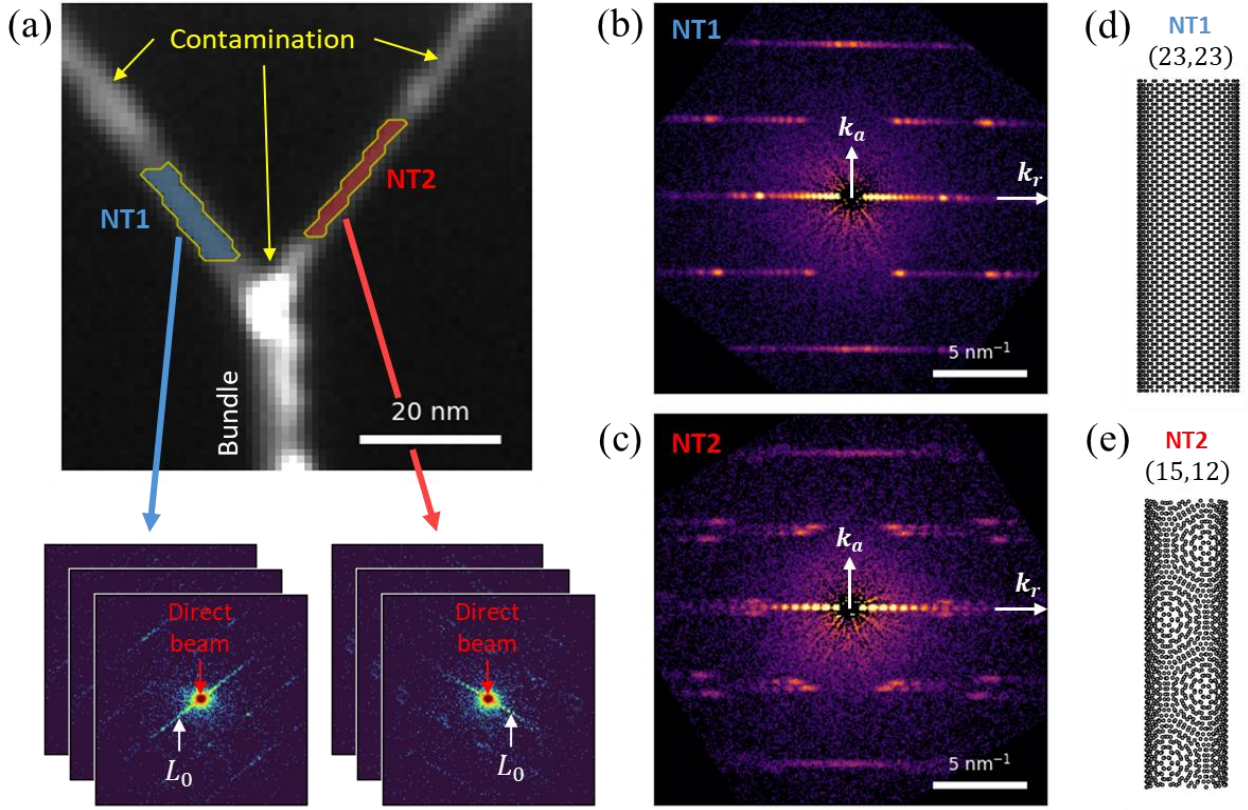

**Figure 2** - 4D-STEM chirality mapping procedure. (a) Virtual ADF image extracted from a 4D-STEM dataset (top), with contaminated areas indicated by the yellow arrows; in the bottom left, stack of diffraction patterns extracted from the area attributed to NT1 (blue area in top) and in the bottom right, stack of diffraction patterns extracted from the area attributed to NT2 (red area in top). (b) and (c) processed PACBEDs associated to NT1 and NT2, respectively. (d) and (e) schematics of the structures determined by the ILLS method for NT1 and NT2, respectively. All PACBEDs are shown without the direct beam and with a square root intensity scale.

To increase the SNR, we can average all patterns in each sub-dataset, resulting in Position Averaged Convergent Beam Electron Diffraction (PACBED) patterns. For a robust averaging procedure, the sub-dataset needs to be descan corrected. When the electron beam is scanned across the sample, it may drift away from the optical axis, leading to an apparent shift of the diffraction patterns on the detector. This is usually compensated for by the descan coils, but some residual drift may still be present and can be numerically corrected by shifting the diffraction pattern images. To do so, we simply measure the position offset of the direct beam with template matching and cross-correlation and realign all diffraction patterns [60]. In addition, we need to compensate for the nanotube curvature. Indeed, nanotubes often curve within the field of view, i.e., the plane perpendicular to the electron beam propagation direction. It results in an apparent rotation of the patterns relative to each other. Curvature compensation is performed by measuring the orientation of the equatorial line, which is always aligned with the tube radial axis and clearly visible in each individual pattern (bottom of **Figure 2**(a)). Note that this requires to first compute a polar transform of the pattern. Then, all images are rotated to align all equatorial lines and all patterns are summed. Finally, we apply a numerical mask on the direct beam and we remove the background intensity using the SNIP algorithm [61].

The resulting pre-processed diffraction pattern are shown in **Figure 2**(b) and **Figure 2**(c), for NT1 and NT2, respectively. They have been rotated so the tube radial direction in reciprocal space, $\boldsymbol{k_r}$, and axial direction, $\boldsymbol{k_a}$, are horizontal and vertical, respectively. In **Figure 2**(c) (which is the same as **Figure 1**b), we can see six layer-lines which means NT2 is a chiral tube. However, **Figure 2**(b) does not contain as many layer-lines and some are superimposed. This is the typical pattern of an armchair tube where $n = m$. In this particular case, we have $G_1 = 0$, $G_2 = G_3 = G_4 = G_5$ and $G_6 = 2G_2$. Note that there is another particular case for zigzag tubes when $m = 0$, which leads to $G_1 = G_2$, $G_3 = 2G_1$, $G_4 = 0$ and $G_5 = G_6 = 3G_1$. The ILLS method is perfectly capable of dealing with these achiral cases [27]. When applied to the PACBEDs in **Figure 2**, it gives a chirality of (23,23) for NT1 and (15,12) for NT2. Simulated diffraction patterns can be found in the supplementary section 1 (Figure S1) for comparison. The corresponding ideal tube structures are represented in **Figure 2**(d) and **Figure 2**(e), for NT1 and NT2, respectively.

Now, the local chirality can be mapped along an individual tube or even multiple tubes, simply by adjusting the size and the location of ROIs and analyzing many PACBEDs or even single diffraction patterns.

### 2.6. *Radial strain mapping*

Strain in carbon nanotubes is known to modify their electronic structure. More interestingly, the radial strain component directly modifies their diameter and affects the nanotube bandgap [8,62]. Most of the time, radial strain relaxes in one direction, leading to an elliptical deformation of the tube [63,64]. When observing the projected diameter of a radially strained SWCNT, it can appear larger or smaller than the unstrained diameter, depending on the projection direction. Therefore, mapping these local projected diameter variations directly provides a radial strain map along the tube. Assuming there is an unstrained reference section with projected diameter $d_{ref}$ on the nanotube, the radial strain component at any location with a projected strained diameter $d_s$ is defined as:

$$\varepsilon_r = \frac{d_s - d_{ref}}{d_{ref}}. \tag{10}$$

With this definition, positive strain values correspond to larger projected diameters, i.e., projections closer to the ellipse's major axis, whereas negative strain values correspond to smaller projected diameters, i.e., projections closer to the minor axis. As discussed in section 2.4, fitting the equatorial line profile allows us to measure diameters. More precisely, it is the tube's diameter projected along the axis perpendicular to the electron beam propagation direction. By applying this fitting procedure at multiple locations along the same tube, we obtain a spatial map of local projected diameter variations. Note that the manual first guess discussed in section 2.4, needs to be passed only once to efficiently measure diameter variations along a single SWCNT. This, in turn, provides the necessary information to construct a radial strain map for an individual nanotube.

Furthermore, the measured radial strain values could be absolute or relative. Determining the absolute radial strain for a specific chirality would require a perfectly calibrated diffraction space and measuring the diameter of an ideal, unstrained nanotube. However, we cannot say for certain that such a nanotube is present in our sample. Instead, we analyze only relative radial strain by selecting a minimally strained reference area and comparing other regions to it. This area is chosen in an uncurved nanotube region and free from residual contamination, providing the closest approximation to an ideal unstrained reference. Although this does not guarantee that the zero radial strain reference corresponds to a perfectly circular tube, it is sufficient to analyze strain variations along the tube and identify the most strained regions.

### 2.7. *Electron ptychography reconstruction*

To generate high-resolution images of carbon nanotubes, we use electron ptychography. It is an advanced phase retrieval technique that reconstructs high-resolution images by analyzing overlapping diffraction patterns collected during a 4D-STEM experiment [65,66]. Unlike conventional imaging methods, ptychography can overcome lens aberrations and extend the resolution beyond the limits of traditional electron optics [38,43,67]. For single-walled carbon nanotubes (SWCNTs), electron ptychography is particularly valuable due to its ability to retrieve high-contrast phase images with sub-angstrom resolution, making it possible to visualize atomic-scale features with minimal electron dose [41].

In this study, simulated and experimental 4D-STEM datasets were processed with the py4DSTEM package [60] to compute ptychographic images. The dataset was first binned down to 128x128 pixels in diffraction space and then reconstructions were carried out using an iterative method based on the stochastic gradient-descent algorithm [68]. We used the mixed-state multi-slice ptychography formalism to account for both the probe decoherence, and the tube diameter (thickness in the beam propagation direction). We used three different probe modes and two slices with a thickness of 3 nm. The reconstructions were additionally constrained using filters both in real and Fourier space to avoid high-frequency artifacts.

## 3. Results & discussion

### *3.1. Chirality mapping*

#### *3.1.1. Chirality determination*

In this section, we discuss the efficiency of the proposed method to determine and map local chiralities in SWCNTs. As shown in **Figure 2**, the chirality mapping procedure allows to determine the chirality of nearby tubes. Here, the chirality of NT1 was measured as (23,23) with $\beta = 4.6°$, NT2 was (15,12) with $\beta = 10.1°$. Prior to the experiment, the diffraction pixel size was calibrated at 0.0814 nm$^{-1}$ using a reference silicon sample. Even though calibration is not mandatory for chirality determination with the ILLS method, it allows us to measure the tube diameters directly from the diffraction patterns. We found a diameter of 31.33 ± 0.17 Å for NT1, while the theoretical diameter for a (23,23) SWCNT is 31.19 Å. Note that the uncertainty in diameter measurement is given with a $3\sigma_{d_0}$ confidence interval (see section 2.4.2). For NT2, the measured diameter is 18.19 ± 0.11 Å, compared to the theoretical diameter of 18.34 Å for a (15,12) SWCNT.

The relative error in the recalculation of $\xi_1^\beta$ for NT2 was 4.4 %, corresponding to an absolute error of 0.16 pixels on the $L_1$ position. As mentioned in section 2.4.2, the error cannot be estimated in the case of NT1 (armchair tube) but the measured diameter is in good agreement with the theoretical value of a (23,23) tube.

Studying this system so close to the bundle separation point would not be possible during a conventional electron diffraction experiment because of the large diameter of the electron beam. Thanks to the capabilities of 4D-STEM, this is now achievable. We now explore the limiting criteria of this technique, particularly the spatial resolution limit, the minimum electron dose for chirality determination, and the influence of the probe convergence angle on the results.

#### *3.1.2. Spatial resolution and minimum dose.*

- **Spatial resolution**

The obvious limiting parameter for spatial resolution is the diameter of the electron beam. In the previous experiment (**Figure 2**), the convergence angle of the probe was measured to be 0.85 mrad which corresponds to a STEM probe full width at half maximum (FWHM) of approximately 2.5 nm. Therefore, the optimal spatial resolution can only be larger than this value. In this case, this is sufficient to characterize the tube structure even within the bundle. **Figure 3** features the same dataset as **Figure 2**, but this time, the dataset is processed to assess both the spatial resolution of the technique and the minimum electron dose required for chirality identification. **Figure 3**(a) is the virtual ADF image where colored pixels show the regions where diffraction patterns were extracted from. The cyan area contains only two patterns corresponding to NT2 and their sum is shown in **Figure 3**(b). The blue, red and green areas contain patterns extracted from the bundle and their respective sums are shown in **Figure 3**(c) (from left to right). We avoided the contaminated area on purpose to investigate only the SWCNT signals. The central red area contains the scattered intensities coming from both tubes, NT1 and NT2 (**Figure 3**(c)). On each side, the left blue area produces the typical pattern of an armchair tube (NT1) while the right green area gives the typical pattern of a chiral tube (NT2). Therefore, the chirality of both tubes can be determined even within the bundle, provided that the beam position is sufficiently far from the line where the tubes are in contact. This minimal distance between adjacent probes to resolved individual tubes in the bundle corresponds to the spatial resolution of the experiment. Here, because the red area has a width of 3 pixels, we can assert that the spatial resolution in this case is 3.1 nm.

To confirm this, we performed the ILLS analysis to determine the chirality of each tubes inside the bundle and compared the results with the previously determined ones. After processing the green patterns in **Figure 3**(c), we measured a chirality of (15,12) for NT2, just as expected. For NT1 (blue pattern in **Figure 3**(c)), we measure a chirality of (20,20) instead of the expected (23,23) indices. This is due to a large radial strain component induced by the bundle in NT1, which is not considered by the ILLS method. We will discuss this aspect later in section 3.2.

Other electron diffraction based techniques such as coherent diffraction imaging (CDI) or ptychography already reached higher spatial resolutions when studying carbon nanotubes [29,41]. However, this is to our knowledge the highest spatial resolution ever reached to characterize SWCNTs chirality purely from electron diffraction, *i.e.*, without phase retrieval data processing.

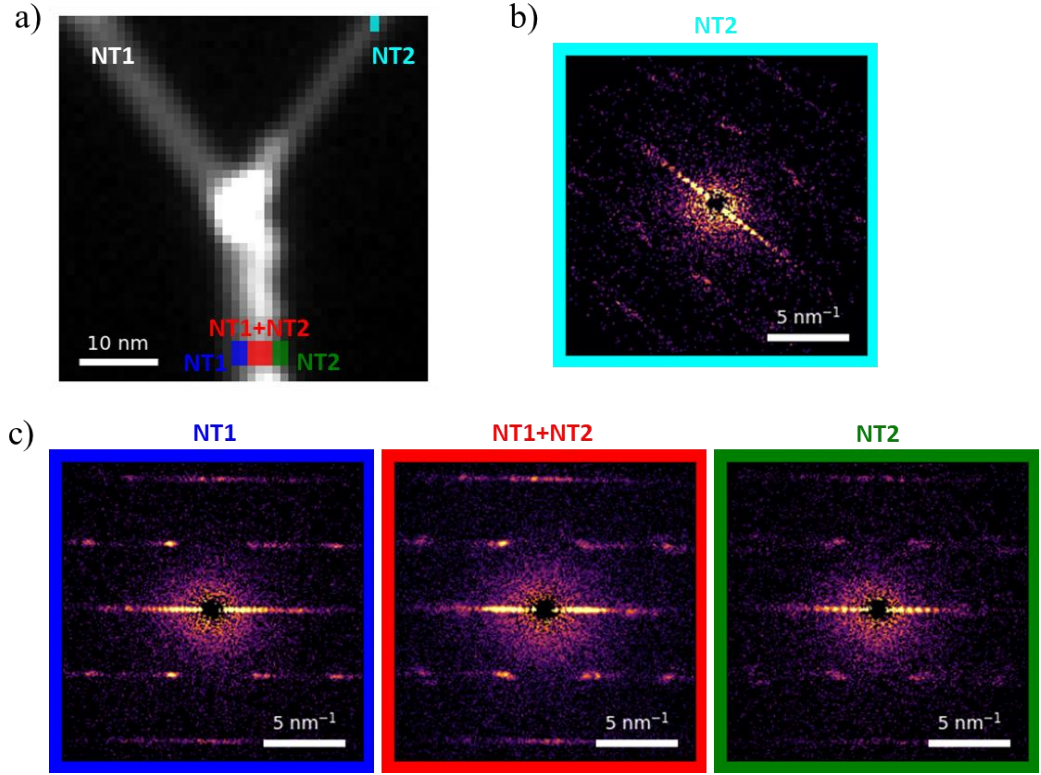

**Figure 3** – Extracted data to assess spatial resolution and minimum electron dose for chirality determination. (a) Virtual ADF image with areas used for minimal dose assessment (cyan) and spatial resolution determination (blue, red and green). (b) PACBED corresponding to cyan area in (a). (c) PACBEDs corresponding to blue, red, and green areas in (a), from left to right. All PACBEDs are shown without the direct beam and with a square root intensity scale.

- **Electron dose**

Then, to determine the minimal electron dose required for chirality determination, we can reduce the number of diffraction patterns within the PACBED and check when the ILLS method fails. With this configuration, we found that we had to sum at least two patterns to determine the chirality of NT2. We performed the ILLS method on the cyan pattern shown in **Figure 3**(b) and measured a chirality of (15,12), just as expected. The diameter was measured at 18.24 ± 0.11 Å, so the reduced electron dose does not affect the confidence interval of the diameter measurement, since $L_0$ maintains sufficient intensity. However, the relative error on the $L_1$ position was 13.9 % (absolute error of 0.59 pixels). Here, the error increases (see section 3.1.1) due to the poor SNR but the chiral indices are still the expected ones. However, the absolute error being more than half a pixel, this is the absolute limit we can reach with our experimental setup. In this experiment, the electron dose per area was estimated at $1.75 \times 10^3$ e$^-$.Å$^{-2}$, so within 2 probe positions the minimal total electron dose is $3.8 \times 10^5$ e$^-$. This lies at the lower end of the typical electron doses used for chirality determination of SWCNTs by electron diffraction [29,30], so our approach does not allow significant dose reduction. However, thanks to the hybrid pixelated detector, it enables working with higher beam currents and shorter exposure times (on the order of tens of milliseconds instead of several seconds), making the acquisition of local chirality maps faster and thus feasible.

### 3.1.3. Probe convergence angle limitations.

As presented in section 2.4.1, to measure the $G_i$ spacings, we measure the peak positions in the integrated intensity profile shown in **Figure 1**(d). Intuitively, one might expect that increasing the probe convergence angle would broaden the peaks. The trivial limitation is that the convergence angle must remain small enough to prevent peak overlap. To define this limiting criterion, we need to express the width of the peak corresponding to one layer line $L_i$, as a function of the convergence angle. Let us formulate the peak intensity profile along $k_a$ as:

$$P_{L_i}(k_a) = \int_{-k_{r_{max}}}^{k_{r_{max}}} I_{L_i}(\mathbf{k}) dk_r, \tag{11}$$

where $I_{L_i}(\mathbf{k})$ is the $L_i$ component of the diffracted intensities and $k_{r_{max}}$ is the maximum recorded spatial frequency on the detector. $I_{L_i}$ is the squared modulus of the complex electron exit wave component associated to $L_i$, $\Psi_{Li}$. Within the kinematical theory of diffraction and using the weak phase approximation, $\Psi_{Li}$ is proportional to the electron probe incident wave function in reciprocal space, $A$, also called the aperture function convolved with the $L_i$ component of the nanotube structure factor, $F_{L_i}$, so

$$\Psi_{Li}(\mathbf{k}) \propto (A \otimes F_{L_i})(\mathbf{k}), \tag{12}$$

where $\otimes$ is the convolution operator [69]. If we neglect all phase errors, i.e., residual probe aberrations[1], $A(\mathbf{k})$ is defined as:

$$A(\mathbf{k}) = \begin{cases} 1, |\mathbf{k}| \leq k_\alpha \\ 0, |\mathbf{k}| > k_\alpha \end{cases}, \tag{13}$$

with $k_\alpha = \frac{\alpha}{\lambda}$, $\lambda$ being the electron wavelength. We can rewrite $P_{L_i}$ using Eqs. (11) and (12) as:

$$P_{L_i}(k_a) = \int_{-k_{r_{max}}}^{k_{r_{max}}} |\Psi_{L_i}(\mathbf{k})|^2 dk_r \propto \int_{-k_{r_{max}}}^{k_{r_{max}}} |(A \otimes F_{L_i})(\mathbf{k})|^2 dk_r. \tag{14}$$

Therefore, the width of the $P_{L_i}$ depends on $F_{L_i}$ and will increase due to the convolution with the aperture function. The broadening is directly related to the size of the aperture; specifically, the larger $k_\alpha$, the wider the peak. After some calculations (see Appendix), we can show that Eq. (14) simplifies to

$$P_{L_i}(k_a) \propto \begin{cases} 4[k_\alpha^2 - (k_a - G_i)^2], & |k_a - G_i| \leq k_\alpha \\ 0, & |k_a - G_i| > k_\alpha \end{cases}. \quad (15)$$

---

[1] Valid in our case because we use a small condenser aperture (5 μm) recovering only the non-aberrated central part of the incident probe.

As expected, $P_{L_i}(k_a)$ reaches its maximum value when $k_a = D_i$, which is proportional to $4k_\alpha^2$. As a result, we can estimate its FWHM which are the distances between the two roots of the equation

$$4[k_\alpha^2 - (k_a - G_i)^2] = 2k_\alpha^2. \quad (16)$$

Solving for Eq. (16), we obtain

$$k_a = G_i \pm \frac{k_\alpha}{\sqrt{2}}, \quad (17)$$

which leads to

$$\text{FWHM} = \sqrt{2}k_\alpha. \quad (18)$$

Finally, to avoid peak overlap, the convergence angle criterion must satisfy the inequality

$$\sqrt{2}k_\alpha < G_i - G_j, \quad (19)$$

$G_i - G_j$ being the distance between two lines $L_i$ and $L_j$.

This criterion can be tested experimentally with close to zigzag, $(n, 1)$, or close to armchair, $(n, n-1)$, tubes. **Figure 4**, shows the 4D-STEM data acquired from a close to armchair SWCNT. The virtual ADF image is presented in **Figure 4**(a), where brighter areas correspond to carbon contamination on the tube and the red areas are the regions used for chirality determination. It was identified as a (14,13) SWCNT with the ILLS method. However, it sits right at the convergence angle criterion limit of the given experimental configuration, requiring further refinement of the data processing protocol to reliably determine its chirality, as detailed in the following. In this experiment the convergence angle was measured to be 0.76 mrad ($k_\alpha = 0.182$ nm$^{-1}$). For a (14,13) SWCNT, $G_1 = G_3 - G_2 = 0.10$ nm$^{-1}$ and $G_5 - G_4 = 0.30$ nm$^{-1}$. The convergence angle criterion previously established gives a minimal resolvable distance between 2 layer-lines of $\sqrt{2}k_\alpha = 0.257$ nm$^{-1}$. Therefore, lines $L_0$ and $L_1$ as well as $L_2$ and $L_3$ are superimposed due to their close spacing. On the other hand, lines $L_4$ and $L_5$ are separated by a distance greater than $\sqrt{2}k_\alpha$ so they should be distinguishable. A simulated diffraction pattern corresponding to a (14,13) SWCNT and an experimental PACBED (extracted from the right most red area in **Figure 4**(a)) are shown in **Figure 4**(b) and **Figure 4**(c), respectively. As expected, lines $L_0$ and $L_1$ as well as $L_2$ and $L_3$ are superimposed, while $L_4$ and $L_5$ are separated in the simulated pattern. In the case of the experimental data (**Figure 4**(c)), this is not as clear, even though the $L_4/L_5$ line intensity is broader than the other lines.

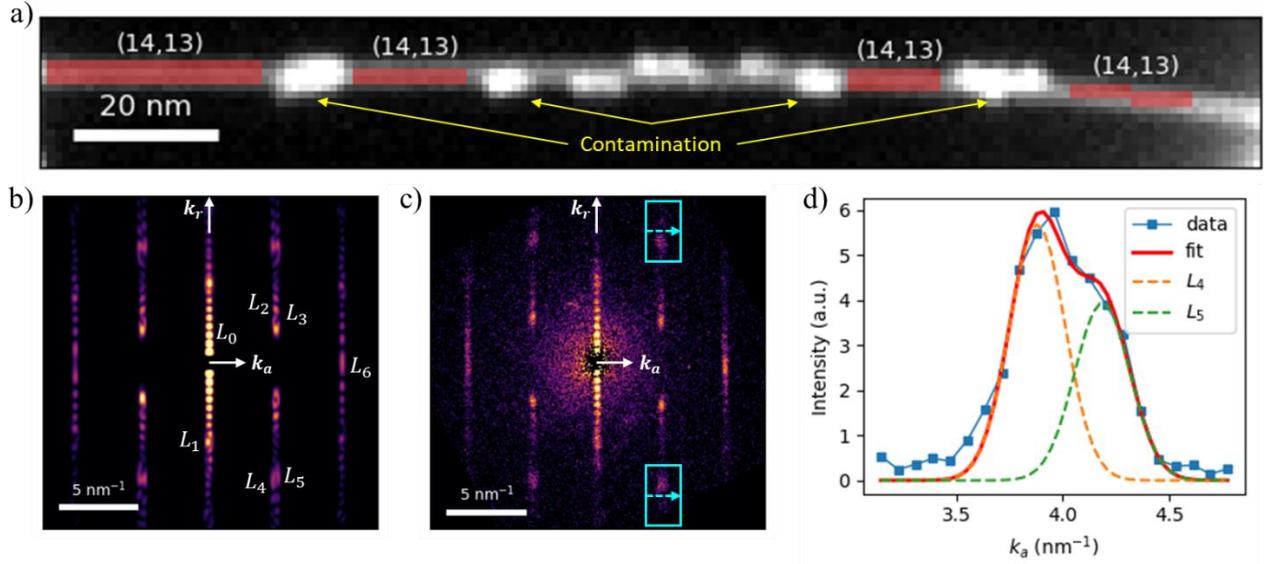

**Figure 4** – Chirality determination of a close to armchair tube with the ILLS + multi-peak fitting method. (a) Virtual ADF image, with contaminated areas indicated by the yellow arrows and in red, areas used for chirality determination. (b) simulated diffraction pattern of a (14,13) SWCNT. (c) PACBED extracted from the right most red area in (a). (d) $k_a$ intensity profiles integrated over $k_r$ extracted from the cyan areas in (c), blue squares connected by a blue line; gaussian fits attributed to the $L_4$ and $L_5$ components, orange and green dashed lines, respectively, and the sum of both fits shown by the red line. All PACBEDs are shown without the direct beam and with a square root intensity scale.

To better assess whether the lines are resolved, we extracted the $P_{L_{4/5}}$ intensity profile from the cyan regions shown in **Figure 4**(c). These profiles are plotted in **Figure 4**(d) as blue squares connected by a blue line. The peak is asymmetric, suggesting that it is composed of multiple components. To model this, two peaks were used to fit the experimental data, corresponding to $L_4$ (orange dashed line in **Figure 4**(d)) and $L_5$ (green dashed line in **Figure 4**(d)). The total fitted curve, shown as a red line in **Figure 4**(d), matches the experimental data well. Although the lines are not perfectly resolved, their individual contributions can still be extracted through multi-peak fitting. The reciprocal space resolution appears lower than in the simulated pattern, probably because of imperfections in the microscope projection system and the detector's modulation transfer function (MTF). Nevertheless, it remains sufficient to measure the peak positions and determine the chirality. Note that instead of fitting with the $P_{L_i}$ functions in Eq. (15), we found that fitting with multiple gaussian functions is more robust to compensate for the detector's MTF. Using the ILLS method combined with multi-peak fitting, all four red regions were consistently identified as having (14,13) chiral indices.

Besides chirality, the diffraction patterns also provide information about strain, allowing for more detailed structural characterization, which is discussed in the next section.

### 3.2. Radial strain mapping

#### 3.2.1. Results

As presented in section 2.6, multiple diameter values can be extracted from equatorial line intensity fits at various locations along the tube, and these can be used to generate radial strain maps. **Figure 5** shows the resulting strain maps obtained from the previously presented datasets. The Y-shape bundle separation strain map is plotted in **Figure 5**(a), superimposed on the virtual ADF image. In addition, **Figure 5**(b) shows the higher resolution HAADF image of the same area. Then, the individual close to armchair nanotube strain map is plotted in **Figure 5**(c), also superimposed on its corresponding virtual ADF image.

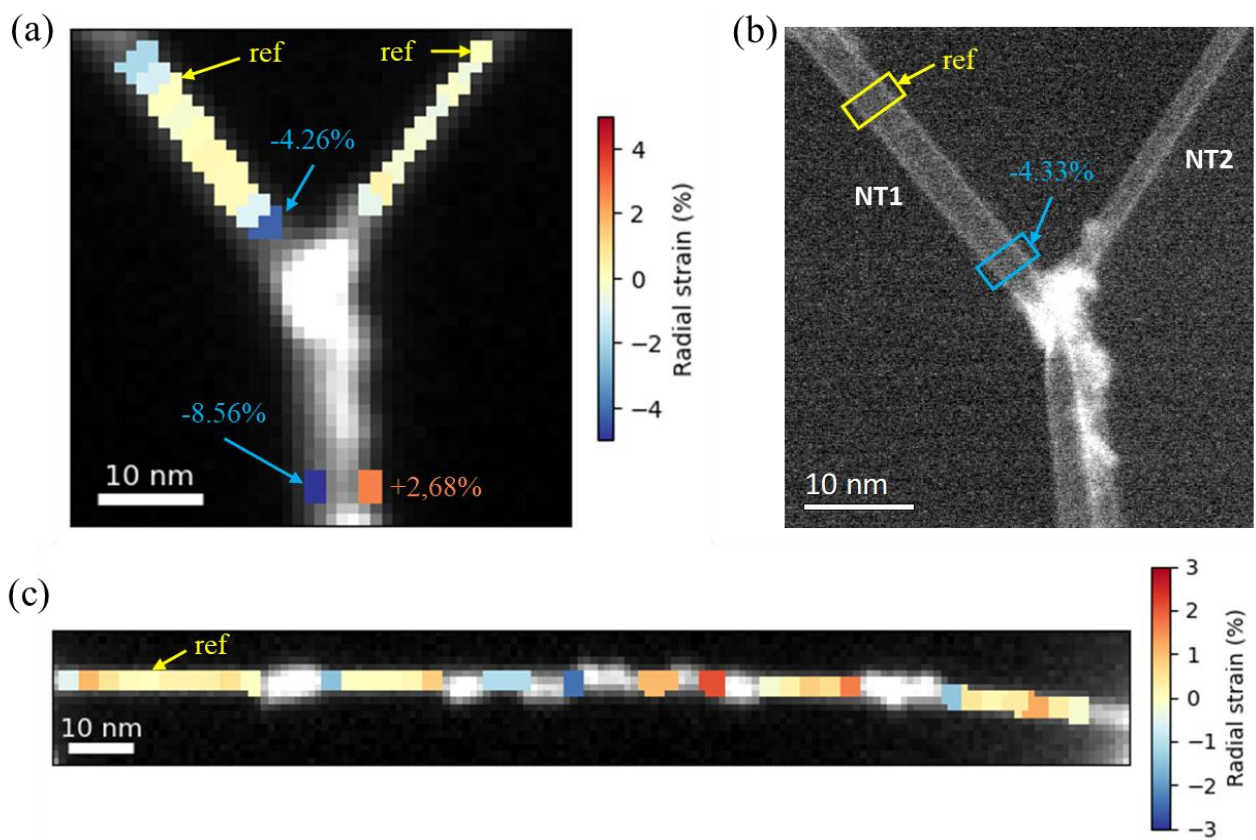

**Figure 5** – Radial strain maps. (a) Virtual ADF image of the Y-shape bundle separation, with superimposed radial strain map. (b) Higher resolution STEM-HAADF of the same area as in (a). (c) Virtual ADF image of the close to armchair tube, with superimposed radial strain map. The reference unstrained areas are indicated by the yellow arrows.

In the Y-shaped bundle separation map (**Figure 5**(a)), the upper region—where the tubes are clearly separated—shows that the nanotubes are mostly unstrained, except for NT1, which exhibits slight negative strain near a contaminated area in the upper left region of **Figure 5**(a)-(b). Closer to the separation point, both nanotubes display negative strain, particularly NT1 at the location indicated by the upper blue arrow in **Figure 5**(a). This strain arises because the nanotubes bend near the separation point, leading to a reduction in projected diameter due to stress relaxation. This buckling behavior in bent SWCNTs is well established in the literature [70,71]. It offers a valuable case for assessing the accuracy of the radial strain measurement method by comparing strain values obtained from diffraction with diameters measured directly in the HAADF image (**Figure 5**(b)). We find good agreement between the two approaches: the manually determined real-space strain is -4.33 %, while the diffraction-based strain is -4.26 %. This level of deformation is fully consistent with the expected buckling of the nanotube [71].

Within the bundle, both nanotubes exhibit significant strain, with opposite trends in projected diameter variation. NT2 is positively strained, showing a projected diameter increase of +2.68 % relative to the reference region, while NT1 is negatively strained with a -8.56 % projected diameter variation. This opposing behavior suggests that both tube experience elliptical deformation in opposite directions, NT1 projecting the minor axis of the ellipse and NT2 the major axis. While van der Waals forces may be responsible for this effect, a definitive conclusion cannot be drawn due to the presence of addition carbon contamination in this region, which can also affect the local tube apparent diameters [63]. Furthermore, the radial strain in NT1 is quite large. Although previous results show no breaking of the armchair symmetry in the diffraction pattern (see **Figure 3**(c)), the observed diameter is much smaller than expected for a (23,23) nanotube. This discrepancy led to a failure in chirality identification using the ILLS method, which incorrectly assigned a nanotube with similar symmetry but smaller diameter (see section 3.1.2). In contrast, NT2 is less strained, so in this case, the ILLS method still correctly identifies the (15,12) chirality. The mean $3\sigma_{d_0}$ confidence interval for the diameter measurement was estimated to be 0.76% here (**Figure 5**(a)).

For the nanotube close to the armchair configuration shown in **Figure 5**(c), the radial strain ranges approximately from −2.4% to 2.1%. Radial strain is primarily observed in contaminated regions, while cleaner areas remain unstrained. This confirms previous results attributing radial deformation to the presence of carbon contamination on SWCNTs [63] and it highlights the importance of nanotube cleanness in maintaining a constant diameter and uniform strain. Such structural stability is critical for nanotube-based device performance, as both diameter and strain directly influence the electronic and transport properties of SWCNTs [15,72–74]. The mean $3\sigma_{d_0}$ confidence interval for the diameter measurement was estimated to be 0.77% in this case (**Figure 5**(c)).

### 3.2.1. *Probe convergence angle limitations.*

The convergence angle also as an impact on radial strain measurements. In the method presented before (section 2.6), the radial strain mapping procedure considers that the equatorial line intensity profile is proportional to the squared modulus of $J_0$ (Eq. (2)). In the case of convergent STEM probe, the diffracted intensities are once again convoluted with the probe aperture function. Therefore, Eq. (2) must be modified to consider the effect of the probe.

$$I_0(\bm{k}) = \left|\Psi_{L_0}(\bm{k})\right|^2 \propto \left|\left(A \otimes J_0(\pi d_0 k_r)\right)(\bm{k})\right|^2. \tag{20}$$

A brute force solution is to directly fit the data with the squared modulus of the convolution scaled by an amplitude factor. This approach is easy to implement since the probe aperture function can be directly measured using a reference vacuum diffraction pattern, assuming that we can ignore phase errors.

Using simulated diffraction patterns and measuring the diameter for a wide range of chiralities (see supplementary section 2), we show that at 80 kV and for convergence angles α > 1 mrad, the residual errors in diameter measurements increase with α when fitting is performed using the simple Bessel function $J_0$ (Eq. (2)). In contrast, the errors remain constant when using the convolution-based model (Eq. (20)). However, the latter model becomes highly non-linear and difficult to fit robustly to experimental data with the currently implemented fitting procedure. For α < 1 mrad, both models yield comparable errors, with $J_0$ even achieving slightly lower residual errors (see supplementary section 2). Based on these observations, we chose to use Eq. (2) for fitting equatorial line profiles and to operate with convergence angles below 1 mrad.

Ideally, the full strain tensor should be determined. However, this requires further investigation to distinguish the effects of axial strain from those of specimen tilt. While measuring shear strain may be feasible, additional study is needed to understand its influence on line positions. These considerations fall outside the scope of the present work.

## 3.3. *Imaging defects with electron ptychography*

### 3.3.1. *Results*

While the previously presented approach (small convergence angle 4D-STEM) provides lots of information about SWCNTs' structural properties, it is still limited by its intrinsic spatial resolution of a few nanometers. This prevents the investigation of atomic scale defects in particular intrinsic topological defects. To overcome this limitation, one approach is to perform 4D-STEM mapping using a larger convergence angle, which improves real-space resolution. This enables the use of electron ptychography to reconstruct phase-contrast images at the atomic scale, allowing direct visualization of atomic structure and local defects in the nanotube.

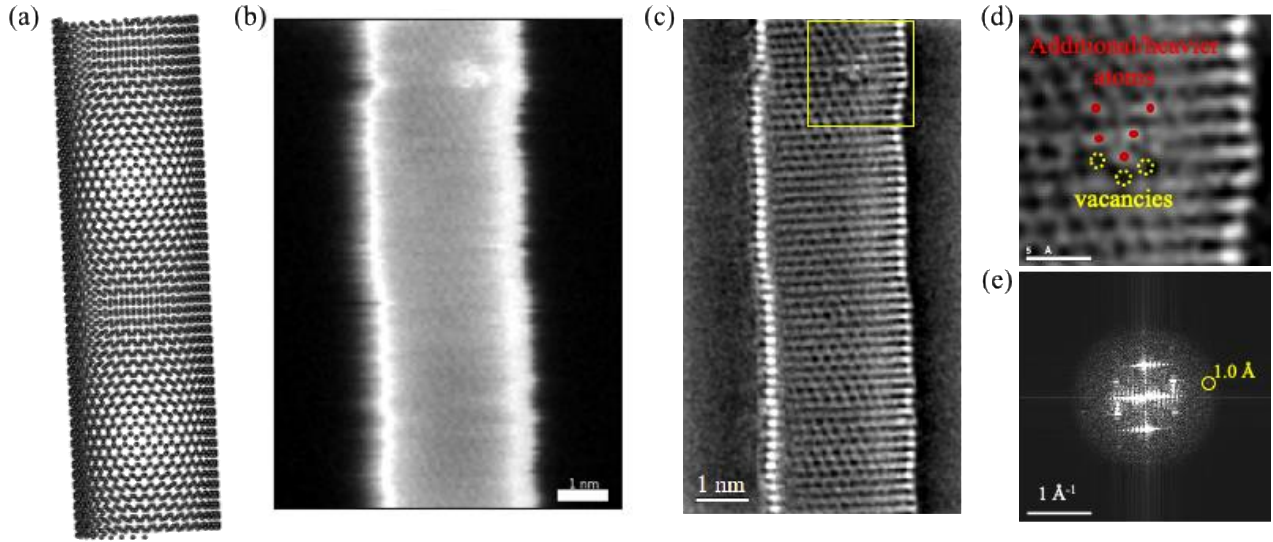

**Figure 6** – Experimental electron ptychography reconstruction of a SWCNT. (a) model of a perfect (37,1) SWCNT. (b) Virtual ADF image of the nanotube. (c) Phase image reconstructed from the same dataset as (b) with electron ptychography. (d) enlarge view of the yellow square in (c). (e) Fourier transform of (c).

**Figure 6** presents the high-resolution images obtained from a 4D-STEM dataset acquired on a defective (37,1) SWCNT. The perfect tube model is shown in **Figure 6**(a) and **Figure 6**(b) shows the virtual ADF image. Then, the phase image of the ptychographic reconstruction is shown in **Figure 6**(c) with an enlarged view of the defective region shown in **Figure 6**(d). Finally, the Fourier transform of the total phase image (**Figure 6**(c)) is shown in **Figure 6**(e).

While the Moiré pattern of the tube and defects is barely visible in the ADF image, it appears clearly in the ptychographic reconstruction. First, the tube chirality can be determined from the phase image by measuring the tube diameter in real space and the $\frac{n}{m}$ ratio from the image FFT (**Figure 6**(e)), following the method detailed in [33]. Here, we determined the tube chiral indices to be (37,1). The FFT enables to determine the spatial resolution of the reconstruction, here measured at 1.0 Å. Finally, because the phase is proportional to the electrostatic potential within the sample, the phase image can be used to analyze the defect structure in the nanotube. It is clearly visible in **Figure 6**(d) that the Moiré pattern symmetry breaks revealing the structure of a topological defect at the surface of the tube. Areas enclosed by yellow dashed circles appear darker than their surroundings, indicating lower potential and suggesting the presence of vacancies. In contrast, areas marked with red dots appear brighter, corresponding to higher potential and possibly representing additional atoms adsorbed on the surface, interstitial carbon atoms, or heavier impurity atoms.

*3.3.2.    Discussion*

Using a large convergence angle in 4D-STEM together with electron ptychography provides improved high-resolution imaging capabilities for studying defects in SWCNTs. Because SWCNTs only scatters weakly the incident electrons, they could in principle be considered as pure thin phase objects [29,75], which require only the single slice ptychography formalism. However, our results suggest that multi-slice ptychography performs better (see supplementary section 3). This is because the wave front of an aberrated probe changes between the front and back sides of the nanotube (see Figure S3 in supplementary section 3), and multi-slice reconstruction helps to compensate for that. For a comparison between mixed-state-only and mixed-state plus multi-slice ptychography reconstructions, see the supplementary section 3 (Figure S4 and Figure S5).

The helical structure of nanotubes produces a moiré pattern in (4D-S)TEM images, arising from the superposition of the tube's front and back sides relative to the electron beam propagation direction. This overlap complicates the interpretation of atomic defect structures. Ideally, separating the two sides would provide deeper insight into the defect structure, which in principle can be achieved using the virtual depth sectioning capability of multi-slice ptychography [43,76]. Here, the idea is to numerically isolate the front and back sides from a single projection. We investigated the theoretical performance of this technique as presented in **Figure 7**. We simulated 4D-STEM datasets obtained from a defective (18,12) nanotube structure extracted from MD simulations. **Figure 7**(a) shows the structure which contains two 5-7 ring pairs, i.e., ensembles of pentagon and heptagon carbon rings instead of the standard hexagon graphene-like arrangement. The tube can be divided into two parts, the frontside, which corresponds to the first part of the tube that encounters the electron beam during propagation, and the backside, which "sees" the electron beam after propagation across the frontside. The 5-7 ring pairs are only present on the backside of the tube, and are marked by blue and green dashed lines in **Figure 7**(a). The goal here is to compare two multi-slice electron ptychography reconstructions, corresponding to two experimental conditions, a first one similar to our experiment, with partially coherent electron beam coming from a X-FEG gun operating at 80 kV (**Figure 7**(b)) and a second one with more coherent electron beam coming from a CFEG gun operating at 200 kV (**Figure 7**(c)).
24

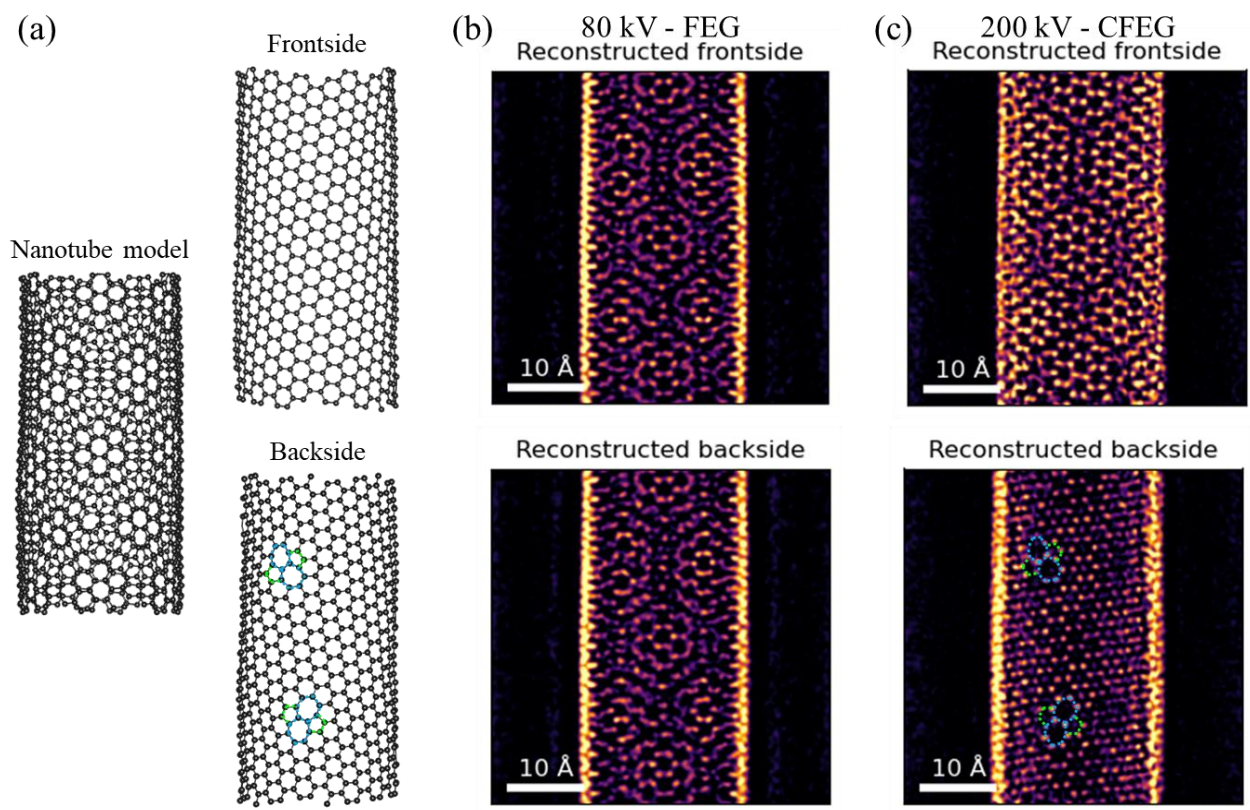

**Figure 7** – Multi-slice electron ptychography reconstructions from simulated 4D-STEM datasets. (a) defective nanotube model projected structure (left), defect free frontside (right top) and defective backside (right bottom). (b) Slices of the multi-slice ptychography reconstruction corresponding to a microscope operating at 80 kV with a X-FEG gun, frontside (top) and backside (bottom). (c) Slices of the multi-slice ptychography reconstruction corresponding to a microscope operating at 200 kV with a CFEG gun, frontside (top) and backside (bottom).

The depth resolution of this technique is limited by the focal spread of the STEM probe which depends on the electron wavelength, convergence angle, and coherence, i.e., energy spread, of the electron beam. The reconstruction from the 80 kV simulated dataset shown in **Figure 7**(b) fails to separate both side of the tube. The tube's moiré pattern is visible in both slices (frontside, top of **Figure 7**(b) and backside bottom of **Figure 7**(b)). This is mainly due to the large focal spread of 25 nm induced by chromatic aberrations in this configuration which is larger than the tube diameter (2 nm). Although the mixed-state formalism can, in principle, compensate for the partial coherence of the electron beam [77,78], its combination with decoherence from thermal vibrations is too significant to allow the ptychographic reconstruction to resolve both sides of the tube. In a more favorable configuration, at 200 kV with a smaller energy spread, the focal spread is now reduced to 3.4 nm. The corresponding mixed-state-multi-slice reconstruction is shown in **Figure 7**(c) and it is able to partially isolate both sides of the tube, even though the focal spread is larger than the tube diameter.

Although the contrast in each slice is not perfect, it is possible to directly visualize both defects (**Figure 7**(c) bottom).

Therefore, multi-slice ptychography could be a way to directly image atomic defects in SWCNTs. However, the approach presented in this study suffers from several flaws. First, due to the small size of SWCNTs and the weak scattering power of individual carbon atoms, it requires large electron doses (on the order of $10^6$ e$^-$·Å$^{-2}$), which can damage the nanotubes during acquisition. Second, achieving sufficient depth resolution to image both sides of a SWCNT from a single projection appears to require operating the microscope at high accelerating voltages (200 kV or higher). Such conditions, however, will induce even stronger knock-on damages during data acquisition and are not expected to be suitable for carbon nanotube imaging. Several technical strategies could, be explored to improve multi-slice ptychography performance on SWCNTs. Possible approaches include increasing the convergence angle and coherence, optimizing the defocus conditions, improving reconstruction algorithm performances, or combining ptychography with tomography. However, assessing the feasibility and efficiency of these strategies would require a dedicated study, which lies beyond the scope of the present paper.

## 4. Conclusion

We have demonstrated the efficiency of 4D-STEM using a hybrid pixelated detector for structural characterization of single-walled carbon nanotubes (SWCNTs) with multiscale capabilities. At small convergence angles, this approach enables efficient chirality determination with spatial resolution down to a few nanometers, making it possible to map chirality variations and track structural changes along the tube. Additionally, strain mapping was successfully implemented, with reliable measurement of the radial strain component. It allows to map projected diameter variations across individual tubes which potentially opens new investigations of the relation between their structural and electronic properties. We also discussed the influence of the STEM probe convergence angle on these measurements. First, they are inherently limited by the trade-off between spatial and reciprocal space resolution, as discussed section 3.1.3. This limitation will become even more significant when analyzing multi-walled carbon nanotubes (MWCNTs), which are more likely to exhibit closed interlayer lines. This will severely restrict the maximum usable probe convergence angle, further reducing spatial resolution. Additionally, radial strain can introduce errors in chirality determination by altering the measured diameter. While Liu and Qin showed that radial strain should not impact chirality determination [63], we argue that large radial deformations can, in fact, lead to incorrect chirality assignment when using the ILLS method. Despite these intrinsic limitations, we were able to define criteria to ensure robust chirality measurement as well as radial strain.

At large convergence angles, 4D-STEM combined with electron ptychography achieves atomic-scale imaging, reaching a spatial resolution of 1.0 Å on SWCNTs. Its strong phase contrast enables both chirality determination as well as visualization of atomic scale defects within nanotubes. Despite this, characterizing defects and distortions remains challenging due to the presence of moiré patterns. While multi-slice electron ptychography improves reconstruction quality, its current numerical depth sectioning capabilities are not sufficient to fully separate both side of the tubes with reasonable electron energies and dose. Even current state-of-the-art imaging techniques could fail to allow direct visualization of atomic defects in SWCNTs. A combination of ptychography and tomography could offer a solution [68], though such experiments remain technically challenging.

Overall, this work paves the way toward multiscale structural characterization of carbon nanotubes, from hundreds of nanometers field of view to atomic-resolution imaging, and could be generalized to other 1D nano-objects.

## 5. Acknowledgements

The authors would like to thank the py4DSTEM team members, Stephanie M. Ribet, Georgios Varnavides, and Colin Ophus for useful discussions regarding ptychographic reconstructions. The authors acknowledge the support of the *Agence Nationale de la Recherche* (ANR) (grant ANR-20-CE09-0007-01). This project received funding from the European Research Council under the European Union's H2020 Research and Innovation program via the e-See project (Grant No. 758385), as well as from the ANR via the PEPR "DIADEM" project METSA (No. ANR-22-PEXD-0013). A CC-BY public copyright license has been applied by the authors to the present document and will be applied to all subsequent versions up to the author accepted manuscript arising from this submission, in accordance with the grant's open access conditions.

## 6. Appendix

This appendix contains mathematical definitions and some demonstrations supporting the paper main text:

- Reciprocal space coordinates:

$$\boldsymbol{k} = k_r \boldsymbol{k_r} + k_a \boldsymbol{k_a} \tag{A1}$$

- $0^{th}$-order Bessel function [26]:

$$J_0(x) = \frac{1}{2\pi}\int_0^{2\pi} e^{ix\cos t} dt \tag{A2}$$

- Layer lines corresponding to $D_i$ ($i = 1,2,\ldots,6$)

$$\begin{aligned}\xi_1 &= \frac{n-m}{\sqrt{3}\pi}, & \xi_2 &= \frac{n+2m}{\sqrt{3}\pi}, & \xi_3 &= \frac{2n+m}{\sqrt{3}\pi},\\ \xi_4 &= \frac{\sqrt{3}m}{\pi}, & \xi_5 &= \frac{\sqrt{3}n}{\pi}, & \xi_6 &= \frac{\sqrt{3}(n+m)}{\pi}.\end{aligned} \tag{A3}$$

The same relations apply for $\xi_i^\beta$ and $(n^\beta, m^\beta)$.

- Squared modulus of the 1D integral of a 2D circular aperture:

Let us define an aperture function in two dimensions of radius $r$ centered at the origin of the coordinates system:

$$A(x,y) = \begin{cases} 1, & x^2+y^2 \leq r^2 \\ 0, & x^2+y^2 > r^2 \end{cases}. \tag{A4}$$

The one-dimensional integral of this function is defined as:

$$I_A(y) = \int_{-\infty}^{+\infty} A(x,y)\, dx. \tag{A5}$$

Considering the function's definition (Eq. A4), Eq. A5 can be written as:

$$I_A(y) = \begin{cases} \int_{-\sqrt{r^2-y^2}}^{\sqrt{r^2-y^2}} dx, & |y| \leq r \\ 0, & |y| > r \end{cases}. \tag{A6}$$

Then, Eq. A6 can be simplified to become:

$$I_A(y) = \begin{cases} 2\sqrt{r^2-y^2}, & |y| \leq r \\ 0, & |y| > r \end{cases}. \tag{A7}$$

This allows to write the squared modulus of the integral of the aperture function which is expressed as:

$$|I_A(y)|^2 = \begin{cases} 4(r^2-y^2), & |y| \leq r \\ 0, & |y| > r \end{cases}. \tag{A8}$$

- Simplification of the $k_a$ peak intensity profile:

Let's simplify the expression of $P_{L_i}(k_a)$ in Eq. (14). To do so, we first need to expand the expression $(A \otimes F_{L_i})(\mathbf{k})$. For any layer-line $L_i$, if we consider only the non-zero $k_r$ component, we know

$$F_{L_i}(k_r) \propto J_\nu(\pi d_0 k_r), \tag{A9}$$

where $J_v$ is a $v^{\text{th}}$-order Bessel function with $v$ a linear combination of the chiral indices $(n, m)$ [26]. Thus, we can write $F_{L_i}(\boldsymbol{k})$ as:

$$F_{L_i}(\boldsymbol{k}) = F_{L_i}(k_r, k_a) \propto [J_v(\pi d_0 k_r, 0) \otimes \delta(0, k_a - G_i)](k_r, k_a), \tag{A10}$$

where $\delta$ is the Dirac delta function. Using the convolution commutativity and the $\delta$ properties with convolution products, we get:

$$\Psi_{L_i}(\boldsymbol{k}) \propto \left(A \otimes F_{L_i}\right)(\boldsymbol{k}) \propto [A(k_r, k_a - G_i) \otimes J_v(\pi d_0 k_r, 0)](\boldsymbol{k}), \tag{A11}$$

Now, Eq. (14) can be written as

$$P_{L_i}(k_a) \propto \int_{-k_{r_{max}}}^{k_{r_{max}}} |[A(k_r, k_a - G_i) \otimes J_v(\pi d_0 k_r, 0)](\boldsymbol{k})|^2 dk_r. \tag{A12}$$

The only term in Eq. (A12) that depends on $k_a$ is $A(k_r, k_a - G_i)$. After all operations: convolution, squared modulus and integration over $k_r$, the final width of $P_{L_i}(k_a)$ depends only on the size of the aperture $k_\alpha$ and on the maximum value of $J_v(\pi d_0 k_r, 0)$. Indeed, convolution with larger absolute values leads to a larger broadening of the line. Note that we are only interested in the proportions of $P_{L_i}$. If we replace $J_v(\pi d_0 k_r, 0)$ in Eq. (A12) with 1, i.e., the maximum value of the normalized function, Eq. (A12) still preserves proportionality. We have

$$P_{L_i}(k_a) \propto \int_{-k_{r_{max}}}^{k_{r_{max}}} |[A(k_r, k_a - G_i) \otimes 1](\boldsymbol{k})|^2 dk_r, \tag{A13}$$

which can also be written as

$$P_{L_i}(k_a) \propto \int_{-k_{r_{max}}}^{k_{r_{max}}} \left| \int_{-\infty}^{+\infty} A(k_r - u_r, k_a - G_i) \, du_r \right|^2 dk_r. \tag{A14}$$

After the first integral, the resulting function inside the squared modulus in Eq. (A14) only depends on $k_a$ which means the integral from $-k_{r_{max}}$ to $k_{r_{max}}$ over $k_r$ only adds another proportionality factor. Finally, Eq. (A14) can be simplified as:

$$P_{L_i}(k_a) \propto \left| \int_{-\infty}^{+\infty} A(k_r - u_r, k_a - G_i) \, du_r \right|^2. \tag{A15}$$

Using Eqs. (A6), (A7) and (A8), we get the following solution for Eq. (A15)

$$P_{L_i}(k_a) \propto \begin{cases} 4[k_\alpha^2 - (k_a - G_i)^2], & |k_a - G_i| \le k_\alpha \\ 0, & |k_a - G_i| > k_\alpha \end{cases}. \tag{A16}$$

# Supplementary to

# Advanced structural characterization of single-walled carbon nanotubes with 4D-STEM

Antonin LOUISET[1], Daniel FÖRSTER[2], Vincent JOURDAIN[3], Saïd TAHIR[3], Nicola VIGANO[1], Christophe BICHARA[4] and Hanako OKUNO[1].

[1]IRIG-MEM, CEA, Université Grenoble Alpes, Grenoble, France.

[2]Interfaces, Confinement, Matériaux et Nanostructures, ICMN, Université d'Orléans, CNRS, Orléans, France.

[3]Laboratoire Charles Coulomb, CNRS, Université de Montpellier, Montpellier, France.

[4]CINaM, CNRS, Université Aix-Marseille, Marseille, France.

*Author to whom correspondence should be addressed: hanako.okuno@cea.fr*

# 1. Experimental and simulated diffraction patterns

Figure S 1 shows the experimental and simulated diffraction patterns discussed in sections 2.5 and 3.1 of the article main text.

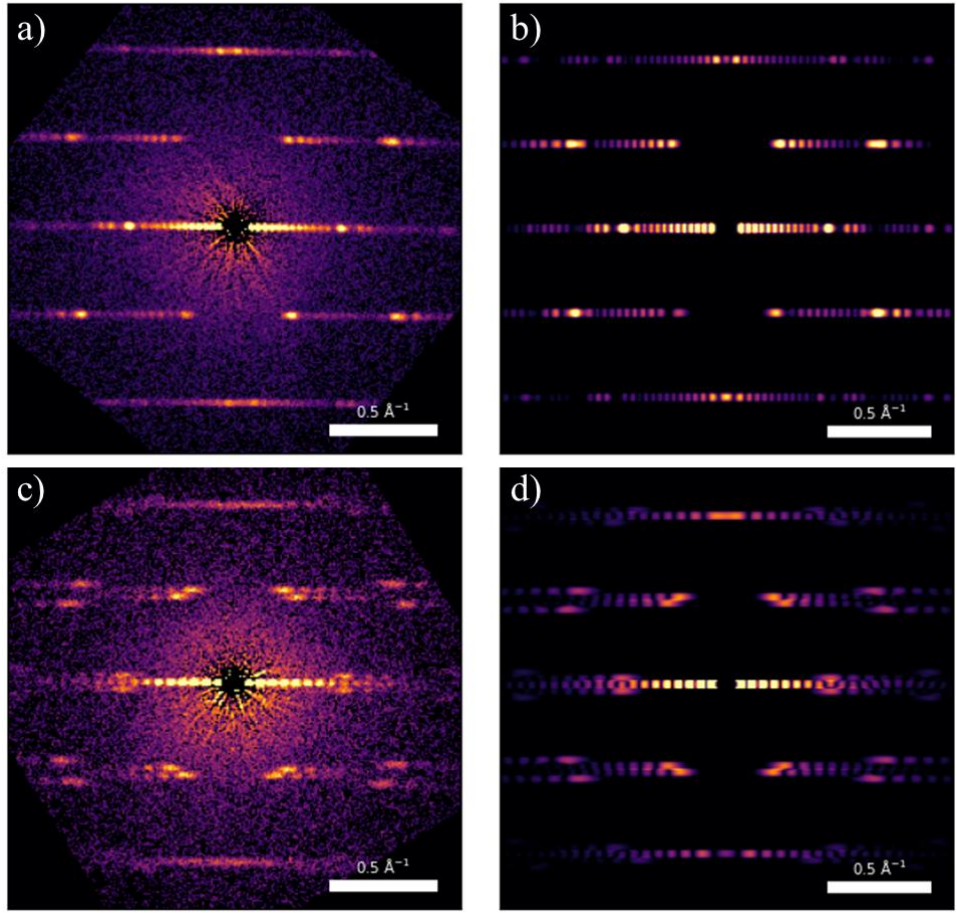

Figure S 1 – Experimental PACBEDs (a, c) and simulated diffraction patterns (b, d) of nanotubes NT1 (a, b) and NT2 (c, d). The direct beam had been masked and images are plotted using a square root intensity scale for better visualization. (b) corresponds to a (23,23) SWCNT tilted at β = 4.6° and (d) to a (15,12) SWCNT tilted at β = 10.1°, as determined by the ILLS method for NT1 and NT2, respectively.

# 2. Convergence angle limitations for radial strain mapping

To evaluate potential errors in diameter measurements, we simulated SWCNT electron diffraction patterns at 80kV for a representative set of chiralities: (15,11), (11,11), (23,17), (14,0), (13,6), (9,4) and (28,3); and for convergence angles between 0.6 and 2.4 mrad. For each diffraction pattern, we determined the tube diameter by fitting the equatorial line intensity profile along $k_r$, using two fitting functions. The first one is the squared modulus of a $0^{th}$-order Bessel function, $|J_0|^2$, which is the standard function used in the literature [1–3] and the second one is the squared modulus of a $0^{th}$ order Bessel function convolved with an aperture function, $|A \otimes J_0|^2$, which we propose as a correction for the convergence angle of the STEM probe. We then calculated the relative diameter errors by comparing the measured diameters with the ground-truth values from the simulation input, using the following relation:

$$d_{error} = \frac{d_{measured} - d_{truth}}{d_{truth}}. \quad (2.1)$$

The results are plotted in Figure S 2. Each marker corresponds to a single chirality and shows the relative error as a function of the probe convergence angle. Blue and orange markers indicate errors obtained when the diameter is measured using the $|J_0|^2$ and the $|A\otimes J_0|^2$ functions, respectively. The averaged errors across all chiralities are shown by the colored lines, with blue corresponding to the $|J_0|^2$ function and orange to the $|A\otimes J_0|^2$ function.

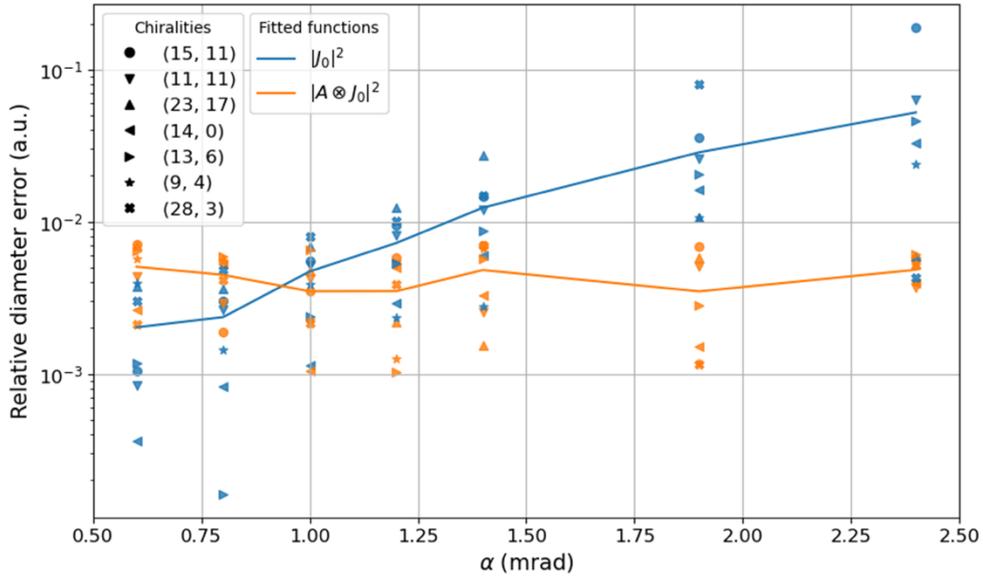

Figure S 2 – Relative errors on diameter measurement from simulated diffraction patterns. Each marker represents one of the 7 chiralities used in the simulations. The data plotted in blue, shows the error on the diameter measurement obtained when fitting the equatorial line profile with the $|J_0|^2$ function, for a given convergence angle. In orange, the same error but using the $|A\otimes J_0|^2$ function to fit the intensity profile is shown. In addition, averaged errors for all chiralities are shown by the colored lines, in blue for $|J_0|^2$ and orange for $|A\otimes J_0|^2$.

For $|A\otimes J_0|^2$, the errors do not increase with the convergence angle and remain within a reasonable range between $10^{-2}$ and $10^{-3}$. This is not the case for $|J_0|^2$, where the errors gradually increase for convergence angles greater than 1 mrad. They reach values of several percents even for angles below 2 mrad which are no longer negligible for strain measurements. Therefore, conducting experiments with larger convergence angles to improve spatial resolution requires correcting for the convolution with the aperture function. However, the $|A\otimes J_0|^2$ model becomes highly non-linear and difficult to fit robustly to experimental data with the currently implemented fitting procedure. For α < 1 mrad, both models yield comparable results, with $|J_0|^2$ even achieving slightly lower residual errors. This is why in this work, we decided to fit the equatorial line profiles with the $|J_0|^2$ function and to operate with convergence angles smaller than 1 mrad.

## 3. Electron ptychography

As introduced in the main text (section 3.3.2), SWCNTs can be described by the weak phase object approximation [4,5], which implies that their atomic structure can be resolved using the single slice ptychography formalism. However, because carbon nanotubes are three-dimensional objects with finite diameters, the wavefront of a convergent electron beam differs slightly between the entrance and exit of a tube. A STEM probe that is in focus on one side of the tube will be out of focus on the opposite side, by an amount corresponding to the tube diameter. For example, in the case of the (37,1) SWCNT presented in this paper, the exit wave front is equivalent to the entrance wave front propagated by 29.8 Å, which corresponds to the theoretical diameter of this tube. This is illustrated in Figure S 3, where the entrance wave front reconstructed with ptychography is shown in Figure S 3(a) and the propagated wave front on the other side of the tube is shown in Figure S 3(b). These images represent electron wave functions by plotting their phase component (color scale) weighted by their amplitude component using color saturation. Because the diameter of the tube is relatively small (only a few nanometers), probe propagation does not drastically affect the wave front. Nonetheless, noticeable differences appear in both the phase and amplitude components, as illustrated in Figure S 3(c). In the following, we discuss how these differences can influence the ptychographic reconstructions.

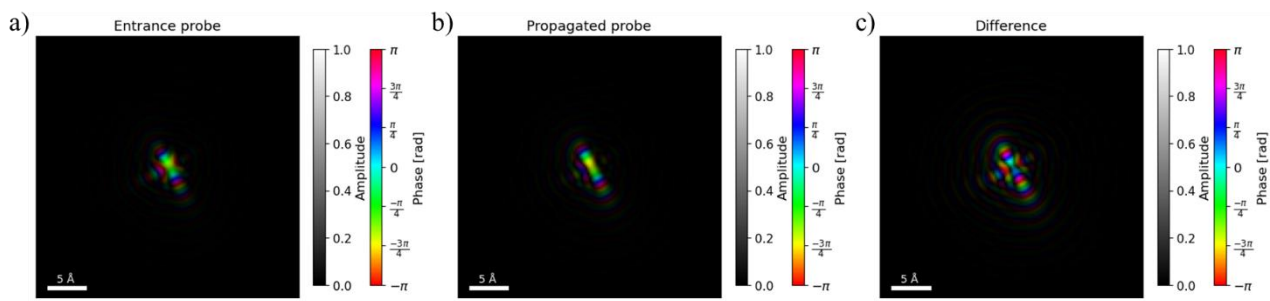

Figure S 3 – (a) Electron probe wave function at the entrance of the nanotube and (b), electron wave function when propagated by 29 Å on the other side of the nanotube. The phase is shown in color scale weighted by the amplitude component using color saturation.

Figure S 4 shows two ptychographic phase images of the (37,1) SWCNT, one reconstructed using the single slice formalism (Figure S 4(a)) and one using the multi-slice formalism with 2 slices of 29 Å (Figure S 4(b), same as in the paper main text, Figure 6). Both reconstructions were run using the same parameters apart for the number of slices, with 3 probe modes. As expected, the phase

images are similar, although the single slice reconstruction exhibits more pronounced artifacts, particularly Fresnel fringes on the left side of the tube.

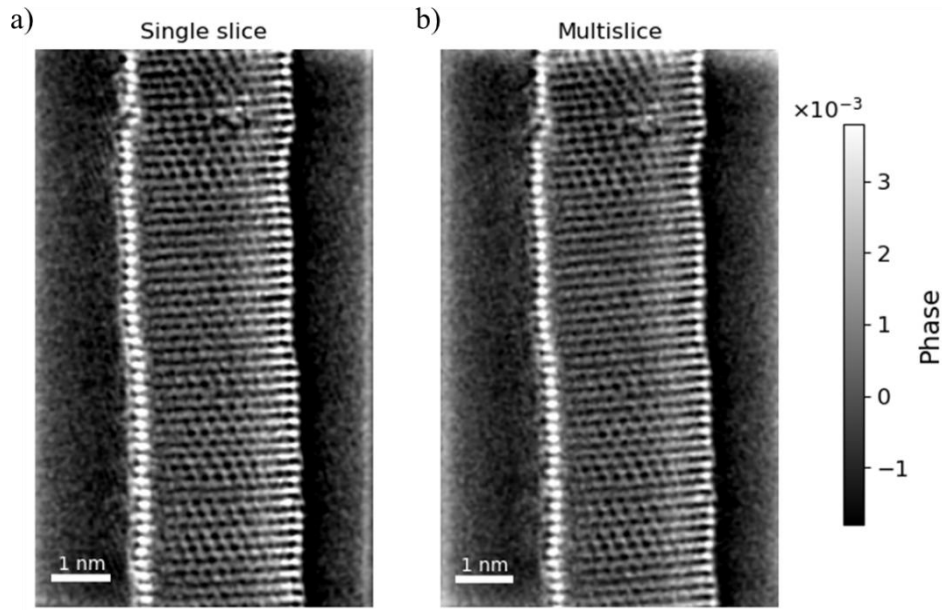

Figure S 4 – (a) Ptychographic reconstruction using the single slice formalism and (b) Ptychographic reconstruction using the multi-slice approach (same as in the main text, **Figure 6**).

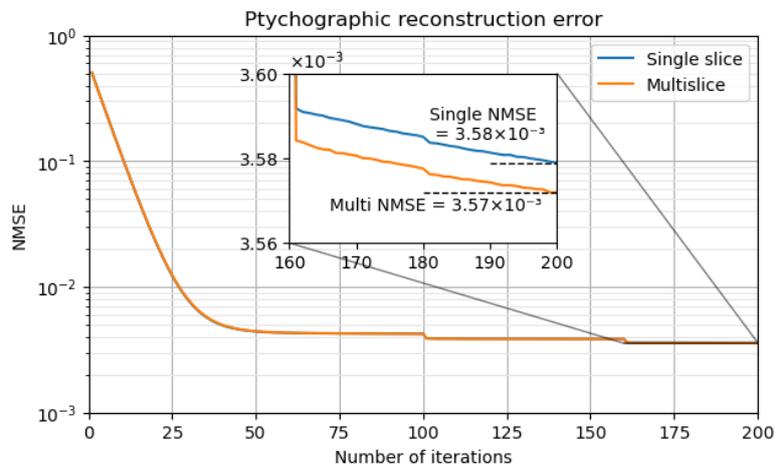

Figure S 5 - Ptychographic reconstruction residual errors as a function of the number of iterations for the single slice reconstruction (blue line) and the multi-slice reconstruction (orange line).

This similarity, with the multi-slice method performing slightly better, is also reflected in the Normalized Mean Squared Error (NMSE) of the reconstructions. The NMSE quantifies the error between the modelled diffracted intensities, computed by applying the object and propagation operators to the modelled probe modes, and the experimental data [6]. Figure S 5 shows the residual NMSE as a function of the number of iterations for both reconstructions. The single slice and multi-slice formalisms are plotted with a blue line and orange line, respectively. The two lines are almost superimposed, indicating comparable performance, however, the multi-slice reconstruction yields lower residual errors, as shown in the inset. Since using only two slices does not significantly increase reconstruction time or computational resource requirements, we adopted the multi-slice formalism for ptychographic imaging in this work.

Finally, Figure S 6 shows the reconstructed probe modes of the multi-slice simulation. All modes are plotted in real (Figure S 6(a)) and reciprocal space (Figure S 6(b)). They are composed of one main probe mode: probe 1, accounting for 98.9% of the total intensity and two additional modes: probe 2 and probe 3, accounting for 0.7% and 0.3% of the total intensity, respectively.

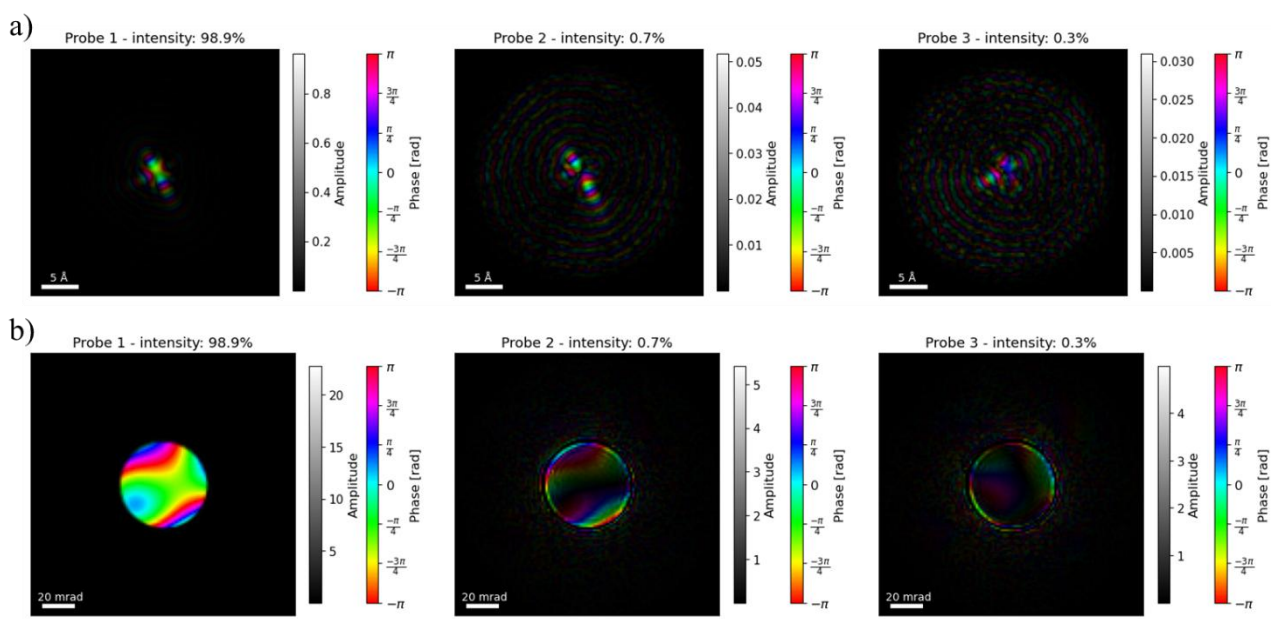

Figure S 6 – Reconstructed probe mode wave functions for the multi-slice reconstruction in real space (a) and in reciprocal space (b).

## 4. Supplementary references